\documentclass[prd,twocolumn]{revtex4}
\usepackage{bm}
\usepackage{epsfig}
\usepackage{amsmath}
\newcommand{\be}{\begin{equation}}
\newcommand{\ee}{\end{equation}}
\newcommand{\ba}{\begin{eqnarray}}
\newcommand{\ea}{\end{eqnarray}}
\newcommand{\bml}{\begin{mathletters}}
\newcommand{\eml}{\end{mathletters}}
\newcommand{\bes}{\begin{subequations}}
\newcommand{\ees}{\end{subequations}}
\newcommand{\ord}{{\cal O}}
\newcommand{\bi}{\begin{itemize}}
\newcommand{\ei}{\end{itemize}}
\newcommand{\gev}{~{\rm GeV}}

\begin{document}
\preprint{UAB-FT-610}
\title{Low-scale inflation in a model of dark energy and dark matter}
\author{P.Q. Hung}
%\email[]{pqh@virginia.edu}
\affiliation{Dept. of Physics, University of Virginia, \\
382 McCormick Road, P. O. Box 400714, Charlottesville, Virginia
22904-4714, USA}
\author{Eduard Mass{\'o}}
%\email{masso@ifae.es}
\author{Gabriel Zsembinszki}
%\email{gabrielz@ifae.es}
\affiliation{Grup de F{\'\i}sica
Te{\`o}rica and Institut de F{\'\i}sica d'Altes
Energies\\Universitat Aut{\`o}noma de Barcelona\\
08193 Bellaterra, Barcelona, Spain}
\date{\today}

\begin{abstract}

We present a complete particle physics model that explains three
major problems of modern cosmology: inflation, dark matter and
dark energy, and also gives a mechanism for leptogenesis. The
model has a new gauge group $SU(2)_Z$ that grows strong at a scale
$\Lambda\sim 10^{-3}$ eV. We focus on the inflationary aspects of
the model. Inflation occurs with a Coleman-Weinberg potential at a
low scale, down to $\sim 6\times 10^5\gev$, being compatible with
observational data.

\end{abstract}



\maketitle

\section{Introduction}

The recent three-year Wilkinson Microwave Anisotropy Probe (WMAP3)
results \cite{wmap3} have put quite a severe constraint on
inflationary models. In particular, new results on the value of
the spectral index $n_{\rm s}=0.95 \pm 0.02$ are sufficiently
``precise'' as to rule out many models with an exact
Harrison-Zel'dovich-Peebles scale-invariant spectrum with $n_{\rm
s}=1$ and for which the tensor-to-scalar ratio $r \ll 1$. Any
model purported to describe the early inflationary era will have
to take into account this constraint. However, by itself, it is
not sufficient to narrow down the various candidate models of
inflation. In particular, ``low-scale'' inflationary models are by
no means ruled out by the new data. By ``low-scale'' we refer to
models in which the scale that characterizes inflation is several
orders of magnitude smaller than a typical Grand Unified Theory
(GUT) scale $\sim 10^{15}- 10^{16}\gev$. It is in this context
that we wish to present a  model of low-scale inflation which
could also describe the dark energy and dark matter \cite{hung},
\cite{hung1}.

The model of dark energy and dark matter described in \cite{hung1}
involves a new gauge group $SU(2)_Z$ which grows strong at a scale
$\Lambda_Z \sim 3 \times 10^{-3}$ eV starting with the value of the gauge
coupling at $\sim 10^{16}\gev$ which is not too different from the
Standard Model (SM) couplings at a similar scale. (This is nicely
seen when we embed $SU(2)_Z$ and the SM in the unified gauge group
$E_6$ \cite{hung2}.) The model of \cite{hung1} contains, in
addition to the usual SM content, particles which are SM singlets
but $SU(2)_Z$ triplets, $\psi^{(Z)}_{(L,R),i}$ with $i=1,2$,
particles which carry quantum numbers of both gauge groups,
$\tilde{\varphi}_{1,2}^{(Z)}$, which are the so-called messenger
fields with the decay of $\tilde{\varphi}_{1}^{(Z)}$ being the
source of SM leptogenesis \cite{hung3}, and a singlet complex
scalar field, \be \label{phiz} \phi_{Z} =
(\sigma_Z+v_{Z})\,\exp(ia_Z/v_{Z})\,, \ee whose angular part $a_Z$
is the axion-like scalar. We have defined the radial part of
$\phi_Z$ as the sum of a field $\sigma_Z$ and a
vacuum-expectation-value (v.e.v.) $v_Z$. The $SU(2)_Z$
instanton-induced potential for $a_Z$ (with two degenerate vacua)
along with a soft-breaking term whose dynamical origin is
discussed in \cite{hung4}, is one that is proposed in \cite{hung1}
as a model for dark energy. In that scenario, the present universe
is assumed to be trapped in a false vacuum of the $a_Z$ potential
with an energy density $\sim \Lambda_Z^4$. The exit time to the
true vacuum was estimated in \cite{hung1} and was found to be
enormous, meaning that our universe will eventually enter a late
inflationary stage.

What might be interesting is the possibility that the real part of
$\phi_{Z}$, namely $\sigma_Z$, could play the role of the inflaton
while the imaginary part, $a_Z$, plays the role of the
``acceleron'' as we have mentioned above. This unified description
is attractive for the simple reason that {\em one complex} field
describes both phenomena: Early and Late inflation. (This scenario
has been exploited earlier \cite{Masso:2006yk}, \cite{others} in
the context of GUT scale inflation.) Although the structure of the
potential describing the accelerating universe is determined, in
the model of \cite{hung1}, by instanton dynamics of the $SU(2)_Z$
gauge interactions \cite{hung1}, the potential for
$\sigma_Z$, which would describe the early inflationary universe,
is arbitrary as with scalar potentials in general. In this case,
the only constraint comes from the requirement that this potential
should be of the type that gives the desired spectral index and
the right amount of inflation corresponding to the characteristic
scale of the model.

We now briefly describe the model of \cite{hung1}. The key
ingredient of that model is the postulate of a new, unbroken gauge
group $SU(2)_Z$ which grows strong at a scale $\Lambda_Z \sim 3
\times 10^{-3}$ eV. The model also contains a global symmetry
$U(1)_{A}^{(Z)}$ which is spontaneously broken by the v.e.v. of
$\phi_{Z}$, namely $\langle \phi_{Z} \rangle = v_Z$, and is also
{\em explicitly} broken at a scale $\Lambda_Z \ll v_Z$ by the
$SU(2)_Z$ gauge anomaly. Because of this, the
pseudo-Nambu-Goldstone boson (PNGB) $a_Z$ acquires a tiny mass as
discussed in \cite{hung1}. Its $SU(2)_Z$ instanton-induced
potential used in the false vacuum scenario for the dark energy is
given by \be \label{totpot} V_{tot}(a_Z,T) =
\Lambda_Z^4\left[1-\kappa(T)\,\cos\frac{a_Z}{v_Z}\right] + \kappa
(T)\Lambda_Z^4\,\frac{a_Z}{2\pi\,v_Z} \,, \ee where $\kappa (T) =
1$ at $T < \Lambda_Z$. ($SU(2)_Z$ instanton effects become
important when $\alpha_Z=g_Z^2/4\,\pi \sim 1$ at $\Lambda_Z \sim 3
\times 10^{-3}\,eV$.) The universe is assumed to be presently
trapped in the false vacuum at $a_Z = 2 \pi v_Z$ with an energy
density $\sim (3 \times 10^{-3}\,eV)^4$. As such, this model
mimicks the $\Lambda$CDM scenario with $w(a_Z)=
\frac{\frac{1}{2}\dot{a}_Z^2 - V(a_Z)} {\frac{1}{2}\dot{a}_Z^2 +
V(a_Z)} \approx -1$, at present and for a long time from now on,
but not in the distant past \cite{hung1}.

What could be the form of the potential for the {\em real part},
namely $\sigma_Z$, of $\phi_{Z}$? As with any scalar field, the
form of the potential is rather arbitrary, with the general
constraints being gauge invariance and renormalizability. In this
paper, we would like to propose a form of potential for $\sigma_Z$
which is particularly suited to the discussion of the
``low-scale'' inflationary scenario: a Coleman-Weinberg (CW) type
of potential \cite{Coleman:1973jx}. (The CW-type of potential has
been recently used \cite{Shafi:2006cs} to describe a GUT-scale
inflation using the WMAP3 data.) There are three types of
contributions to the potential. The sources of these three types
are the following terms in the lagrangian:

\bi

\item[a)] $\phi_{Z}-\psi^{(Z)}_{(L,R),i}$ coupling

\be \label{coupling1} \sum_{i} K_{i} \,
\bar{\psi}^{(Z)}_{L,i}\,\psi^{(Z)}_{R,i}\,\phi_{Z} + h.c. \ee Let
us recall from \cite{hung1} that (\ref{coupling1}) is invariant
under the following global $U(1)_{A}^{(Z)}$ symmetry
transformations: $\psi^{(Z)}_{L,i} \rightarrow
e^{-i\alpha}\,\psi^{(Z)}_{L,i}$, $\psi^{(Z)}_{R,i} \rightarrow
e^{i\alpha}\,\psi^{(Z)}_{R,i}$, and $\phi_{Z} \rightarrow
e^{-2i\alpha}\,\phi_{Z}$.

\item[b)] $\phi_Z - \tilde\varphi_1^Z$ mixing (we ignore
the $\phi_Z - \tilde\varphi_2^Z$ coupling since it is
assumed to have a mass of the order of a typical GUT scale)
\be\label{coupling2}
\tilde \lambda_{1Z}(\phi_Z^{\dag}\,\phi_Z)(
\tilde\varphi_1^{Z,\dag}\,\tilde\varphi_1^Z)\ee

\item[c)] $\sigma_Z$ self-interaction \be\label{coupling3}
\frac{\lambda}{4!} \sigma_Z^4\ee Both terms, (\ref{coupling2}) and
(\ref{coupling3}), arise from the general potential for all
fields.

\ei Let us look into constraints on these couplings coming from
issues discussed in \cite{hung1}: dark matter and leptogenesis.

Since the coupling (\ref{coupling1}) will, in principle,
contribute to the CW potential for $\sigma_Z$, it is crucial to
have an estimate on the magnitude of the Yukawa couplings $K_{i}$.
In \cite{hung1}, an argument was made as to why it might be
possible that $\psi^{(Z)}_{i}$ could be Cold Dark Matter (CDM)
provided \be \label{cdm} m_{\psi^{(Z)}_{i}}= |K_i| v_Z \leq
O(200\,GeV)\,, \ee or \be \label{Ki} |K_i| \leq O(200\,GeV/v_Z)
\ee Roughly speaking, in order for $\Omega_{CDM} \sim O(1)$, the
annihilation cross sections for $\psi^{(Z)}_{i}$ are required to
be of the order of weak cross sections. In this case, they are
approximately $\sigma_{\rm annihilation} \sim \alpha_Z^2(
m_{\psi^{(Z)}_{i}})/m_{\psi^{(Z)}_{i}}^2$ ($\alpha_Z^2(
m_{\psi^{(Z)}_{i}})$ is the coupling evaluated at $E =
m_{\psi^{(Z)}_{i}}$) and have the desired magnitude when
$m_{\psi^{(Z)}_{i}} \sim O(200\,GeV)$ with
$\alpha_Z^2(m_{\psi^{(Z)}_{i}}) \sim \alpha_{SU(2)_L}^2
(m_{\psi^{(Z)}_{i}})$, a characteristic feature of the model of
\cite{hung1}.

A second requirement comes from a new mechanism for leptogenesis
as briefly mentioned in \cite{hung1} and described in detail in
\cite{hung3}. This new scenario of leptogenesis involves the decay
of a messenger scalar field, $\tilde{\varphi}_{1}^{(Z)}$, into
$\psi^{(Z)}_{i}$ and a SM lepton. In order to give the correct
estimate for the net lepton number, a bound on the mass of
$\tilde{\varphi}_{1}^{(Z)}$ was derived. In \cite{hung3}, it was
found that \be \label{lepto} m_{\tilde{\varphi}_{1}^{(Z)}} \alt
1\,TeV \,. \ee
This came about when one calculates the interference
between the tree-level and one-loop contributions to the decays
\be
\label{decay1}
\tilde{\varphi}_{1}^{(Z)} \rightarrow \bar{\psi}^{(Z)}_{1,2}+ l
\ee
\be
\label{decay2}
\tilde{\varphi}_{1}^{(Z),*} \rightarrow \psi^{(Z)}_{1,2}+ \bar{l}
\ee
where $l$ represents a SM lepton. By requiring that the
asymmetry coming from this scenario to be
$\epsilon^{\tilde{\varphi}_{1}}_{l} \sim -10^{-7}$
in order to obtain the right amount of baryon number
asymmetry through the electroweak sphaleron process, \cite{hung3}
came up with the constraint (\ref{lepto}) which could
be interesting for searches of non-SM scalars at
the Large Hadron Collider (LHC).
On the other hand, as discussed in \cite{hung1},
the mixing between $\tilde{\varphi}_{1}^{(Z)}$ and $\phi_{Z}$
results in an additional term in the mass squared formula for
$\tilde{\varphi}_{1}^{(Z)}$, namely $2
\tilde{\lambda}_{1Z}v_{Z}^2$. Taking into account the leptogenesis
bound (\ref{lepto}), one can write \be \label{cross}
\tilde{\lambda}_{1Z} \alt (1\,TeV/v_Z)^2 \,. \ee

All these constraints will be used to estimate the contributions
to the effective potential. The $\psi^{(Z)}_{i}$ fermion loop
contribution to the $\sigma_Z$ CW potential is given by
$-|K_i|^4/16\pi^2$, and therefore will be bound by (\ref{Ki}). The
$\tilde{\varphi}_{1}^{(Z)}$ loop contribution to the potential is
given by $\tilde{\lambda}_{1Z}^2/16\pi^2$ and is constrained by
(\ref{cross}). The third contribution c) coming from the
$\sigma_Z$ loop is given by $\lambda^2/16\pi^2$. There are no
constraints on it coming from dark matter or leptogenesis
arguments, as we have in the other cases a) and b).

Below, we will constrain both $v_Z$ and the coefficient of the CW
potential (which includes contributions from various loops) using
the latest WMAP3 data. Next, we use these results to further
constrain $|K_i|$ and $\tilde{\lambda}_{1Z}$. We will finally
comment on the implications of these constraints.

\section{Inflation with a Coleman-Weinberg potential}

Let us now see under which conditions we can obtain a viable
scenario for inflation with our model. As previously mentioned,
the scalar field $\phi_Z$ receives various contributions to its
potential, which will have the generic CW form
\cite{Coleman:1973jx}

\be
V_0(\phi_Z^\dagger\phi_Z)=A\left(\phi_Z^\dagger\phi_Z\right)^2\left(
\log{\frac{\phi_Z^\dagger\phi_Z} {v_Z^2}}-\frac12\right)
+\frac{Av_Z^4}{2}. \label{cw-potential}\ee After making the
replacement (\ref{phiz}) in (\ref{cw-potential}), we obtain the
potential for the real part $\sigma_Z$ of $\phi_Z$, which we want
to be the inflaton field \be V_0(\sigma_Z)=A(\sigma_Z+v_Z)^4\left[
\log{\frac{(\sigma_Z+v_Z)^2}
{v_Z^2}}-\frac12\right]+\frac{Av_Z^4}{2}. \label{cw-potential2}
\ee This expression corresponds to the zero temperature limit. If
we take into consideration finite-temperature effects, we should
add a new term depending on temperature $T$ that will give the
following effective potential to $\sigma_Z$ \be V_{\rm
eff}(\sigma_Z) = V_0(\sigma_Z) +\beta T^2(\sigma_Z+v_Z)^2
\label{cw-effective} \ee where $\beta$ is a numerical constant. At
high temperature, the field $\sigma_Z$ is trapped at the
$U(1)_A^{(Z)}-$symmetric minimum $\sigma_Z=-v_{Z}$. As the
universe cools, for a sufficiently low temperature a new minimum
appears at the $U(1)_A^{(Z)}-$symmetry breaking value $\sigma_Z
=0$ ($\langle \phi_{Z} \rangle = v_Z$). The critical temperature
is the temperature at which the two minima become degenerate and
is equal to $T_{\rm cr}=v_Z\sqrt{A/\beta}\;e^{-1/4}$. The universe
cools further with the field $\sigma_Z$ being trapped at the false
vacuum and inflation starts when the false vacuum energy of
$\sigma_Z$ becomes dominant. Nevertheless, when the universe
reaches the Hawking temperature
 \be
   T_H=\frac{H}{2\pi}\simeq\frac1{2\pi}\sqrt{\frac{8\pi}{3M_{\rm P}^2}
   V_0(-v_Z)}=\frac{\sqrt{A}\;v_Z^2}{\sqrt{3\pi}\;M_{\rm P}}\label{T_cr}
 \ee
a first-order phase transition occurs and $\sigma_Z$ may start its
slow-rolling towards the true minimum of the potential. In
(\ref{T_cr}), $M_{\rm P}\simeq 1.22\times 10^{19}\gev$ is the
Planck-mass, $H$ is the Hubble parameter at that epoch and we
supposed that the energy density of $\sigma_Z$ is the dominant
one. Observable inflation occurs just after the false vacuum is
destabilized and the inflaton slowly rolls down the potential. The
evolution of $\sigma_Z$ can be described classically.

Next, let us find the values for $A$ and $v_Z$ that are needed in
order to obtain a viable model for inflation, compatible with
observational data. The main constraints come from the combined
observations of the Cosmic Microwave Background (CMB) and the
Large Scale Structure (LSS) of the universe, which indicate the
range of values for the spectral index $n_{\rm s}$, the
tensor-to-scalar ratio $r$ and, perhaps, evidence for a running in
the spectral index. We also consider the constraint on the
amplitude of the curvature perturbations, ${\cal P_R}^{1/2}$, with
the assumption that they were produced by quantum fluctuations of
the inflaton field when the present large scales of the universe
left the horizon during inflation. Finally, the number of e-folds
of inflation produced between that epoch and the end of the
inflationary stage should be large enough in order to solve the
horizon and the flatness problems.

In our scenario, we have a theoretically motivated mechanism for
generating a lepton asymmetry which then translates into a baryon
asymmetry compatible with observations \cite{hung1}, \cite{hung3}.
Later on in this paper we will treat this aspect in more detail.
For now, it is sufficient to say that after inflation, the
$\sigma_Z$ field starts to oscillate and to reheat the universe,
mainly by decaying into two $\psi_i^{(Z)}$ fermions of masses
given by (\ref{cdm}), so that we want the inflaton to have
sufficient mass as to decay into the two fermions. This is another
condition to be considered for obtaining the adequate values for
the parameters of our model.

Let us list the main constraints to be imposed on our model: \bi
\item the spectral index \be n_{\rm s}\simeq 1-6\epsilon+2\eta \ee
should be in the range $n_{\rm s}=0.95 \pm 0.02$, where $\epsilon=
\frac{M_{\rm P}^2}{16\pi}\left(\frac{V^\prime}{V}\right)^2$ and
$\eta= \frac{M_{\rm P}^2}{8\pi}\frac{V^{\prime\prime}}{V}$ are the
slow-roll parameters and a prime means $\sigma_Z$-derivative;
\item the right number of e-folds of inflation between large scale
horizon crossing and the end of inflation \be
N_0=\int_{\sigma_{Z,end}}^{\sigma_{Z,0}} \frac{V}{V'}d\sigma_Z \ee
where $\sigma_{Z,0}$ is the value of the inflaton field at horizon
crossing and $\sigma_{Z,end}$ its value at the end of inflation;
\item the amplitude of the curvature perturbations generated by
the inflaton, evaluated at $\sigma_{Z,0}$ \be {\cal
P_R}^{1/2}=\sqrt{\frac{128\pi}{3}}\frac{V^{3/2}}{M_{\rm P}^3
|V^{\prime}|}|_{\sigma_{Z,0}}\ee should have the WMAP3 value
${\cal P_R}^{1/2}\simeq 4.7\times 10^{-5}$; \item the inflaton
mass $m_{\sigma_Z}=\sqrt{8A}\,v_Z$ should be at least $400\gev$ or
so in order for $\psi_i^{(Z)}$'s to be produced by the inflaton
decay.\ei

In our analysis, the parameters are functions of the inflaton
field $\sigma_Z$ and are evaluated when the present horizon scales
left the inflationary horizon.

We have performed a complete numerical study. Imposing the
requirements we have mentioned, we are interested in the lowest
possible scale for inflation in our model. The scale is lowest for
$v_Z\simeq 3\times 10^9 \gev$. With this value, we obtain $A\simeq
3\times 10^{-15}$ and $m_{\sigma_Z}\simeq 450 \gev$ which, as
commented above, is sufficient to produce two $\psi_i^{(Z)}$
fermions of masses $\ord (200)\gev$. We also obtain $n_{\rm
s}=0.923$ for the spectral index, not far away from the observed
range, and $N\simeq 38$ e-folds of inflation between the present
large scales horizon crossing until the end of inflation. The
inflation scale is $V_0^{1/4}\equiv V(\sigma_{Z,0})^{1/4} \simeq
6\times 10^5\gev$. Low-scale inflation is interesting because it
might be proved more easily in particle physics experiments.

\begin{figure}[htb]
%\begin{center}
\includegraphics[width=8cm, height=6cm]{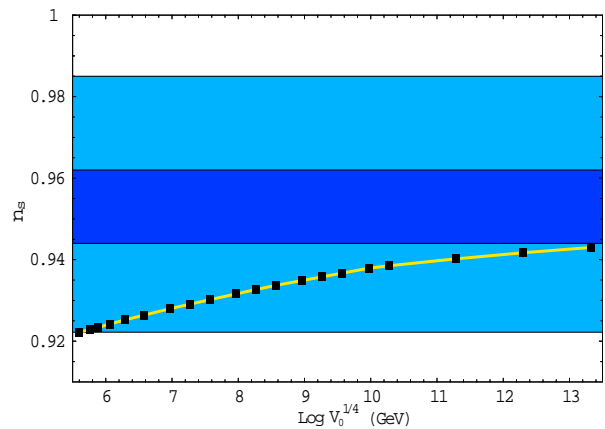}
%\end{center}
\caption{The spectral index $n_{\rm s}$ as a function of the
logarithm of the scale of inflation, $\log V_0^{1/4}$, compared
with WMAP3 \cite{wmap3} range for $n_{\rm s}$ (at 68\% and 95\%
confidence levels)}\label{spectral_index}
\end{figure}

Our model also satisfies the constraints for values of the
parameters leading to higher values than $V_0^{1/4}\simeq 6\times
10^5\gev$. In Fig. \ref{spectral_index} we show the dependence of
the spectral index {\it vs} the energy scale of inflation. We see
that the values of $n_{\rm s}$ are within 95\% confidence level
even at scales as low as $6\times 10^5\gev$, and increases with
increasing inflationary scale. The graphic displayed in Fig.
\ref{spectral_index} was obtained with the assumption of instant
reheating and a standard thermal history of the universe
\cite{inflation}. In Section \ref{reheating} we present a more
detailed analysis on the reheating mechanism. Here we just mention
that the reheating temperature in our model is smaller than
$V_0^{1/4}$, in which case the values of $n_{\rm s}$ displayed in
Fig.\ref{spectral_index} shift to smaller values.

From now on we will stick to the lowest possible example
$v_Z=3\times 10^9\gev$ $(V_0^{1/4}\sim 6\times 10^5\gev)$ and
examine the consequences. We should stress that the values of $A$
does not vary drastically when we raise $V_0^{1/4}$ and we can
safely consider it constant, with the value $A\simeq 3\times
10^{-15}$. We then study some of the consequences that arise when
adopting this value for $A$.

As stated before, one can have one-loop contributions to the
parameter $A$ coming from loops containing a) fermions
$\psi_i^{(Z)}$, b) the messenger field $\tilde\varphi_1^{(Z)}$,
and c) the inflaton. The fermion loop contribution, of order
$-|K_i|^4/16\pi^2$, can be estimated for the values of the
parameters chosen in the previous numerical example. From
(\ref{Ki}) we get for $v_Z=3\times 10^9\gev$ \be |K_i|\lesssim
6.7\times 10^{-8}\ee which will then translate into the following
contribution to the $A$ parameter in the CW potential \be
A_{\psi}\approx - |K_i|^4/16\pi^2\sim 10^{-31} \ee obviously being
too small to be considered as contributing to it. Thus, fermion
loops are completely negligible.

Next, we want to estimate what the contribution of the messenger
field $\tilde\varphi_1^{(Z)}$ is. The leptogenesis bound
(\ref{cross}) for $v_Z=3\times 10^9\gev$ becomes \be
\tilde\lambda_{1Z}\lesssim 10^{-13} \ee which gives the following
contribution to the potential \be
A_{\tilde\varphi}\approx\tilde\lambda_{1Z}^2/16\pi^2 \sim 8\times
10^{-29}\ee also being too small compared to the value $A=3\times
10^{-15}$. This means that the main contribution should come from
$\sigma_Z$ self-coupling $\lambda$. The necessary value of the
$\lambda$ coupling can be estimated by comparing its contribution
$A_\sigma \approx\lambda^2/16\pi^2$ with $A$ \be A\simeq
A_\sigma\approx \lambda^2/16\pi^2 \simeq 3\times 10^{-15} \ee from
which we obtain the constraint on $\lambda$ \be \lambda \simeq
6.9\times 10^{-7}. \ee

To end the discussion regarding inflation in our model, we would
like to add that in our numerical study we obtained a small value
for the running of the power spectrum, $\alpha\equiv\frac{{\rm
d}n_{\rm s}}{{\rm d}\ln k}\simeq -0.002$. Other parameter that
might be of interest is the tensor-to-scalar ratio $r$, which is
defined usually as \be r=\frac{{\cal P}_T}{{\cal P}_R}\ee where
${\cal P}_T$ and ${\cal P}_R$ are the power spectra for tensor and
scalar perturbations, respectively. In the slow-roll regime of
inflation, $r$ can be expressed in terms of the slow-roll
parameters and, at first order, $r=16\epsilon$, where $\epsilon$
has to be evaluated at horizon crossing. With the values used in
our previous numerical example, we obtain a very small
tensor-to-scalar ratio $r\sim 10^{-43}$, making the quest for
gravitational wave detection from the inflationary epoch hopeless.

It is amusing to note that the value of the $\sigma_Z$
self-coupling $\lambda \sim O(10^{-7})$ that is consistent with
the data is of the same order as the constraint on the Yukawa
coupling $|K_i|$ coming from the CDM scenario of \cite{hung1}.

\section{Reheating}\label{reheating}

One of the most important questions of any inflationary scenario
is the following: How do SM particles get generated at the end of
inflation? In a generic inflationary model, it is not easy to
answer this question since a generic inflaton is usually not
coupled, either directly or indirectly, to SM particles. Although
our inflaton is a SM-singlet field, we will show that its decay
and the subsequent thermalization of the decay products can
generate SM particles. In what follows, we will assume that
the inflaton decays perturbatively as with the ``old''
reheating scenario and study its consequences. The interesting question
of whether or not it can decay through the parametric resonance
mechanism \cite{parametric} of ``preheating'' scenarios is beyond
the scope of this paper and will be dealt with separately elsewhere.

\begin{figure}[htb]
%\begin{center}
\includegraphics[width=8cm, height=5.5cm]{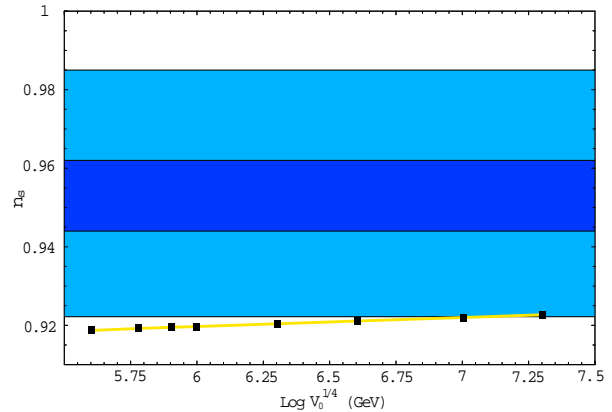}
%\end{center}
\caption{The spectral index $n_{\rm s}$ as a function of the
logarithm of the inflationary scale $V_0^{1/4}$, compared with
WMAP3 \cite{wmap3} range for $n_{\rm s}$ (at 68\% and 95\%
confidence levels), in the range allowed after imposing
constraints coming from the reheating mechanism
}\label{spectral_index2}
\end{figure}

At the end of inflation, the inflaton will rapidly roll down its
potential to the true minimum. The reheating (or, equivalently,
the damping of the inflaton oscillation) occurs via the decay \be
\label{decay} \sigma_Z \rightarrow \psi^{(Z)} + \bar{\psi}^{(Z)}.
\ee The width of the decay (\ref{decay}) is given by \be
\Gamma(\sigma_Z\rightarrow 2\psi^{(Z)})\simeq
9\left(\frac{m_{\psi}}{v_Z}\right)^2\frac{m_{\sigma}}{8\pi}\beta
\label{width1} \ee where
$\beta=(1-4(m_{\psi}/m_{\sigma})^2)^{3/2}$ and we remember that
$m_{\sigma}=\sqrt{8A}\,v_Z$. To estimate the reheating temperature
caused by the process (\ref{decay}) after the end of inflation, we
write \be \Gamma(\sigma_Z\rightarrow 2\psi^{(Z)})\sim H_{\rm
rh}\label{width2} \ee where $H_{\rm rh}\sim 1.66 T_{\rm
rh}^2/M_{\rm P}$ is the Hubble parameter at the reheating
temperature $T_{\rm rh}$. By combining Eqs. (\ref{width1}) and
(\ref{width2}) we obtain the dependence of the reheating
temperature $T_{\rm rh}$ on $v_Z$ \be T_{\rm rh}\simeq 1.3\times
10^8 \left(\frac{v_Z}{\gev}\right)^{-1/2}. \label{T_rh}\ee We see
that the reheating temperature is a decreasing function of $v_Z$.
This will set an upper bound on $v_Z$, because $T_{\rm rh}$ should
be larger than twice the mass of $\psi^{(Z)}$ in order for the
reheating mechanism to work, i.e. \be T_{\rm rh}>2m_{\psi}\sim
400\gev \ee which combined with (\ref{T_rh}) gives \be
v_Z<10^{11}\gev. \ee This upper limit restrict us to a low-scale
inflation range, $6\times 10^5\gev \lesssim V_0^{1/4} \lesssim
2\times 10^7\gev$, and then great part of Fig.
\ref{spectral_index} will be excluded, unless some other reheating
mechanism is invoked. The spectral index values as a function of
the logarithm of the scale of inflation, in the allowed range, is
shown in Fig. \ref{spectral_index2}. Notice that the values of
$n_{\rm s}$ are a bit smaller now than in the case of instant
reheating, but still marginally compatible with the WMAP3 value
for $n_{\rm s}$.

Let us focus now on the mechanism by which SM particles are
produced. For the sake of clarity in the following discussion, we
will denote the QCD gluons by $\tilde{g}$ and the $SU(2)_Z$
"gluons" by ${\bf G}$. The chain of reactions which finally leads
to the SM particles can be seen as follows: \be \label{reheating1}
\psi^{(Z)} + \bar{\psi}^{(Z)} \rightarrow {\bf G}\,{\bf G}
\rightarrow \tilde{\varphi}_{1}^{(Z)}\,\tilde{\varphi}_{1}^{(Z)}
\rightarrow W\, W ,\,Z\, Z \rightarrow q\,\bar{q},\, l \,
\bar{l}\,, \ee and \be \label{reheating2} q\,\bar{q} \rightarrow
\tilde{g}\tilde{g}\,. \ee We end up with a thermal bath of SM and
$SU(2)_Z$ particles. This thermalization is possible because of
the simple fact that $\tilde{\varphi}_{1}^{(Z)}$ carries {\em both
SM and $SU(2)_Z$ quantum numbers}. Another important point
concerns the various reactions rate in
(\ref{reheating1},\ref{reheating2}). The corresponding amplitudes
are proportional to $O(g^2)$, where $g$ stands for either the
$SU(2)_Z$ coupling or a typical SM coupling at an energy above the
electroweak scale. From \cite{hung1} and \cite{hung2}, it can be
seen that the various gauge couplings are of the same order of
magnitude for a large range of energy, from a typical GUT scale
down to the electroweak scale. One can safely conclude that the
various reaction rates are comparable in magnitudes and the
thermalization process shown above is truly effective. In
principle, the messenger field also couples to $\psi^{(Z)}$ and a
SM lepton, as shown in \cite{hung1}, but this is irrelevant in the
thermalization process because the corresponding Yukawa couplings
are too small.

It is remarkable to notice also that, because of the quantum
numbers of the messenger field, the decay of
$\tilde{\varphi}_{1}^{(Z)}$ into a SM lepton and $\psi^{(Z)}$ can
generate a net SM lepton number which is subsequently
transmogrified into a net baryon number through the electroweak
sphaleron process as shown in \cite{hung3}. In other words, the
crucial presence of the messenger field
$\tilde{\varphi}_{1}^{(Z)}$ facilitates both the generation of SM
particles through thermalization and the subsequent leptogenesis
through its decay.

\section{Conclusions}

In this paper, we show that the model presented in \cite{hung1},
which explains dark matter and dark energy, also provides a
mechanism for inflation in the early universe. We find that it is
conceivable to have a low-scale inflation.

The complete model contains a new gauge group $SU(2)_Z$ which
grows strong at a scale $\Lambda\sim 3\times 10^{-3}$ eV, with the
gauge coupling at GUT scale comparable to the SM couplings at the
same scale. In addition to the SM particles, the model contains
new particles: $\psi_{(L,R),i}^{(Z)} (i=1,2)$ which are $SU(2)_Z$
triplets and SM singlets; $\tilde\varphi_{1,2}^{(Z)}$ which are
the so-called messenger fields and carry charges of both $SU(2)_Z$
and SM groups; and $\phi_Z$, which is a singlet complex scalar
field. The model also contains a new global symmetry
$U(1)_A^{(Z)}$, which is spontaneously broken by the v.e.v. of the
scalar field, $\langle\phi_Z\rangle=v_Z$, and also explicitly
broken at the scale $\Lambda_Z\ll v_Z$ by the $SU(2)_Z$ gauge
anomaly. The real part of the complex scalar field, namely
$\sigma_Z$, is identified with the inflaton field. We considered a
CW-type of potential for $\sigma_Z$ and obtained the constraints
on the parameters of the model in order to have a right
description of inflation. The angular part of the complex scalar
field, namely $a_Z$, acquires a small mass due to the explicit
breaking of $SU(2)_Z$ and is trapped in a false vacuum, being
responsible for the dark energy of the universe. The new particles
$\psi_{(L,R),i}^{(Z)}$, with masses of order $200\gev$, explain
the dark matter. They are produced at reheating, by the decay of
the inflaton. For values of the $SU(2)_Z$ breaking scale $v_Z\sim
3\times 10^9\gev$, we obtain a low-scale model of inflation,
namely a scale of $\sim 6\times 10^5\gev$. Notice that,
in order to have a realistic reheating mechanism, this ``low-scale''
is also bounded from above by $\sim 2\times 10^7\gev$ as we have
discussed in the last section. Because of this fact, our model
is a bona-fide ``low-scale'' inflationary scenario.
It is an exciting
possibility because the model might be indirectly probed at
future LHC experiments.

\begin{acknowledgments}
We would like to thank Qaisar Shafi for communicating a graph
similar to Fig. \ref{spectral_index}. The work of PQH is supported
in parts by the US Department of Energy under grant No.
DE-A505-89ER40518. PQH wishes to thank Lia Pancheri, Gino Isidori
and the Spring Institute for the hospitality in the Theory Group
at LNF, Frascati, where part of this work was carried out. The
work of EM and GZ is supported by Spanish grant  FPA-2005-05904
and by Catalan DURSI grant 2005SGR00916. GZ is also supported by
DURSI under grant 2003FI00138.
\end{acknowledgments}

\end{document}